# Understanding training in superconducting accelerator magnets using acoustic emission diagnostics

Maxim Marchevsky

*Accelerator Technology and Applied Physics Division, Lawrence Berkeley National Laboratory*

*mmartchevskii@lbl.gov*

**Abstract**

Among various long-standing standing problems in superconducting accelerator magnet development, training is one of the oldest and well-known issues affecting magnet performance. Understanding physics behind the training process is crucial for proper mitigation and elimination of this phenomenon in new magnet designs. This paper discusses application of acoustic emission (AE) diagnostics to localization and classification of transient mechanical events in superconducting magnets and characterization of transient energy releases responsible for premature quenching, memory effects and, ultimately, training. Prospective of future AE diagnostics development are outlined in connection to the goals of US Magnet Development Program.

**Introduction**

The phenomenon of "training" has baffled superconducting magnet designers for over 40 years [1-7]. Training is a process of gradually improving magnet performance with repetitive quenches required to reach the design current. It is a costly and time consuming procedure that nearly every newly constructed magnet has to undergo prior to its intended use. A striking example is the commissioning and training campaign of 1232 dipoles of the LHC that has consumed several years of magnet test facility operations [8]. Upon thermal cycling magnets would typically sustain the "memory" of the previously reached current level, but eventual de-training may also occur [9]. Finding causes of training and eliminating them would dramatically reduce costs and shorten development time for future colliders. High-temperature superconductors (HTS) are being increasingly advertised as a training-free alternative to the present technology [10] but given their high manufacturing costs and uncertainty with respect to stress-tolerance [11] and protection [12] it is likely that either $Nb_3Sn$ or $Nb_3Sn$/HTS hybrids will be used in next generation accelerators and so the training issue will remain standing for a foreseeable future. Certain empirical recommendations have been developed early on by the superconducting magnet community with respect to training reduction [13,14], and successful implementation of those recommendations allowed to accelerate or even completely eliminate training for some magnets [15]. However, no *universal* solution has ever been found, and every newly built magnet still represents an unknown with respect to its training behavior. With the present proposal we aim at using advanced instrumentation to address the physics of training and seek a universal solution to this problem that would be independent of a particular magnet design.

It can be argued that training is such a hard problem to solve because our means of studying its underlying physics remain very limited. Also, there are significant unknown factors complicating a meaningful analysis. The first such factor is the elastic energy $U_{el}$ stored in the magnet at a given time. It is proportional to the accumulated strain ($\sim \varepsilon^2$), and depends upon structural and material aspects of magnet fabrication, amount of pre-stress, number of cooling cycles and energization history. Magnets are mechanically non-conservative: $U_{el}$ is partially converted into heat during energization thought slip-stick conductor motion or cracking of the impregnation material, and training can be seen as a process that gradually "clears" this non-conservative behavior. Since for untrained magnets stress and strain distributions are not uniquely related, a "stress management" approach recently adopted by high-field magnet designers [16,17] was only occasionally successful in reducing training. Secondly, the exact nature of transient energy releases remains elusive, and it unclear if a particular quench triggering event is in any way "special" compared to other mechanical transients (and therefore can be identified as a "quench precursor"). Cracking and slip-stick motion occur deep inside the magnet coil windings and only take a fraction of a millisecond in most cases. This makes them inaccessible to the majority of existing diagnostic techniques. Finally, it is unclear if the mechanical memory is a statistical phenomenon or a result of some self-organizing behavior, and if any macroscopic parameters can be measured in the magnet to evaluate such memory without actually going through the quenching process.

**Acoustic emission diagnostics**

Addressing these standing questions is crucial for understanding training. We intend to do so using acoustic diagnostic methods [18,19] that are uniquely suitable for this purpose. They allow for a direct yet non-invasive monitoring of mechanical transients, structural integrity and thermal releases in the bulk of the magnet. Acoustic instrumentation is robust and inexpensive compared to other diagnostics (such as fiber-optics) that are under development to gain similar capabilities. Acoustic methods were shown to:

  i.   Spatially localize mechanical transients and quench locations using time of flight techniques [20,21]
  ii.  Measure local energy release in those transients by relying on the known calibration methods [22]
  iii. Record transient acoustic waveforms with a high temporal resolution and apply advanced spectral analysis techniques based on deep learning to classify various mechanical event types

Our diagnostics development involves AE sensor hardware, data acquisition and data analysis.

In Fig 1 cryogenic AE sensors are shown that were recently developed at LBNL for the US Magnet Development Program (US MDP). Newer sensor design will rely upon latest semiconductor parts operating at cryogenic temperatures and the "waveguiding" principle where acoustic signals can be picked up from some specific and hard to access point inside the magnet thus improving signal fidelity and localization capabilities.

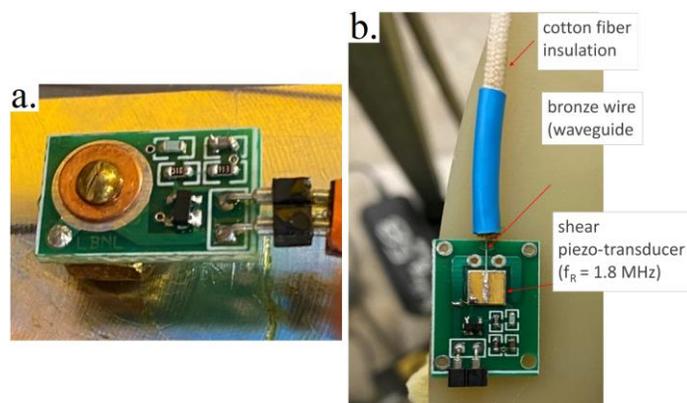

Fig. 1. (a) An AE sensor with integrated cryogenic amplifier having ~ 300 kHz of useful frequency bandwidth, mountable directly on the magnet surface. (b) A new design of AE sensor having a shear-type piezoelectric element installed directly on the amplifier board and acoustic signals fed to from a remote point using an insulated metal wire – acoustic "waveguide". The sensor has an extended frequency bandwidth of > 1 MHz and provides higher fidelity of the AE signal.

Improved localization of transient sources of mechanical energy release relies on a multi-step processing of bare acoustic emission signals, including de-noising, de-trending and multi-level thresholding to achieve best localization accuracy. Fig. 2 shows an example of 3D quench localization in Canted Cosine Theta $Nb_3Sn$ superconducting magnet CCT4 build by the US MDP [17]. In this case an array of 8 AE sensors was installed on the magnet outer shell and quench locations determined by triangulation with pre-processed AEs associated with the quench onset. Accuracy of ~ 5 cm can be assumed based on the data acquisition rate (1 MHz) and sensor frequency bandwidth. Accuracy is expected to improve when a larger number of sensors is used and if acoustic waveguides are employed to pick up signals from additional structural locations.

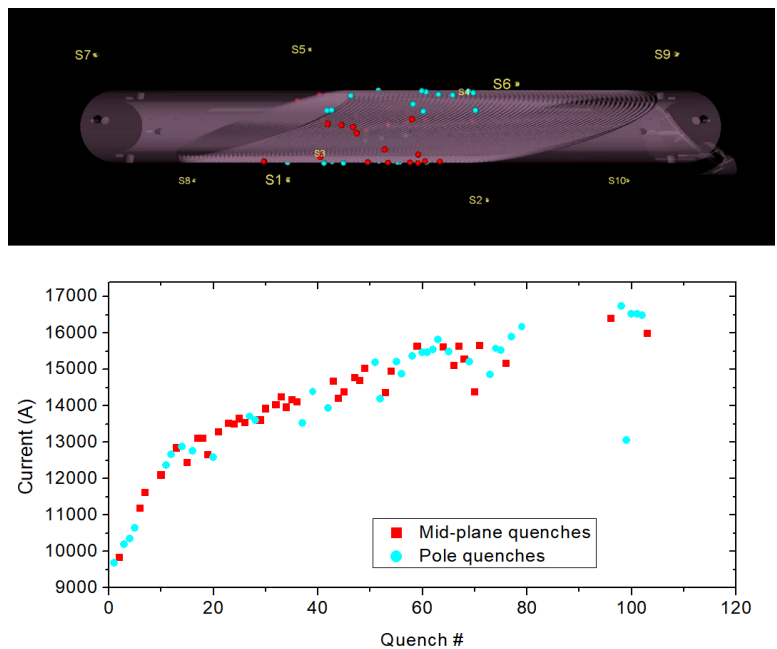

Fig. 2 (top) A map of quench location obtained using AE triangulation, superimposed with a 3D map of the CCT4 coil. (bottom) The training plot of same magnet.

Mechanical energy release is one of the key quantities that can be accesses with AE diagnostics. The square of AE voltage is proportional to the mechanical energy, and calibration can be done by applying external mechanical excitations [21]. An interesting evolution can be observed when average energy release per event is plotted versus the magnet current. In the process of training, a transition occurs from a regime with no apparent scaling into a regime with clear universal power-law scaling that breaks down just prior to the quench. This result suggests that critical dynamics emerges in the magnet at a certain current value and is sustained throughout the current ramp until the quench. Also, a step divergence from the power-law developing a few seconds prior the quench for some ramps suggests presence of acoustic precursors to imminent quenching. This interesting behavior needs to be further understood.

When a total acoustic energy release over current ramp to quench is plotted as function of quench number, a dependence can be observed that is characterized by initial reduction of net energy release with every subsequent training quench of the magnet, until some minimum is reached. That minimum often coincides with change in slope of the training curve. With continuing training, the net energy release starts to gradually increasing again reaching towards some universal asymptotic slope. An example of this behavior in the subscale magnet CCT Sub4 is shown in Fig. 4. One can speculate that this observed dependence of AE energy versus quench number can be explained by the initial creation of defects (cracks, de-laminations) in the magnet coils followed by a transition to slip-stick motion regime. However, more direct evidence is needed to confirm such interpretation of the data.

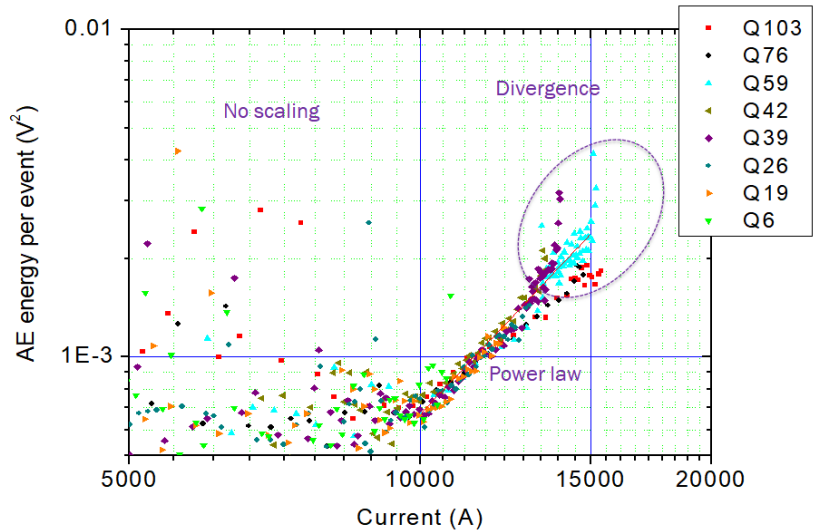

Fig. 3. Acoustic energy per event plotted for several training ramps of the CCT4 magnet.

Classification of mechanical transients by type (cracking, de-lamination, slip-stick, etc.) is another very important task for AE data analysis. Machine Learning (ML) methods, and in particular unsupervised learning algorithms seem to be most useful for such analysis. In the example shown below (Fig. 5) all transient AE events acquired in the training process of CCT4 were separated from the acquired data stream, amplitude-normalized and analyzed using discreet wavelet transforms (6-level Daubechies db2 DWT). Scale coefficients related to event frequency content were found, and a file containing "fingerprint" of raw acoustic data was generated for every ramp. Two fingerprint files, of quench #1 and quench #103 were then labeled as type"1" and "type" 2" respectively. The labeled tata was randomized and split into two equally-sized subsets of 2860 events each. The subsets were used to train Random Forest Classifier and verify its ability to satisfactory distinguish events coming from quench #1 and quench #103. Next, the trained classifier was applied to other quenches in the training set, and the fraction of type "1" events (= similar to events of quench #1) was plotted versus the quench number, as shown in Fig. 5. The plot reveals a similarity with the magnet training curve, also showing a steep change in the fraction of type "1" events where the quench training plot exhibits a significant change in slope. Work should be continued to develop better classifiers for transient events and collect more AE data statistics to better train our ML models.

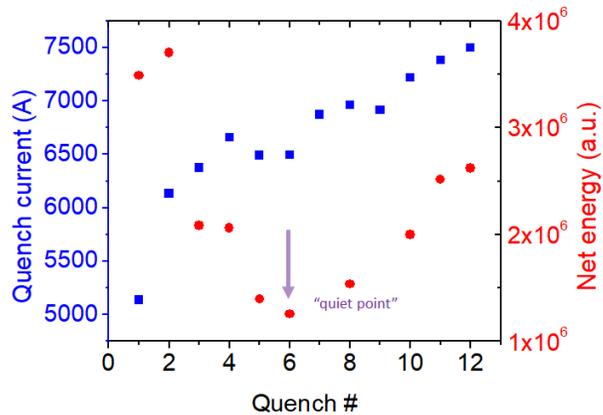

Fig. 4. Net AE energy release plotted versus quench number, superimposed with the magnet training curve for the subscale CCT Sub4. The minimal energy release is reached by quench #6 and then gradually increases again with continuing training.

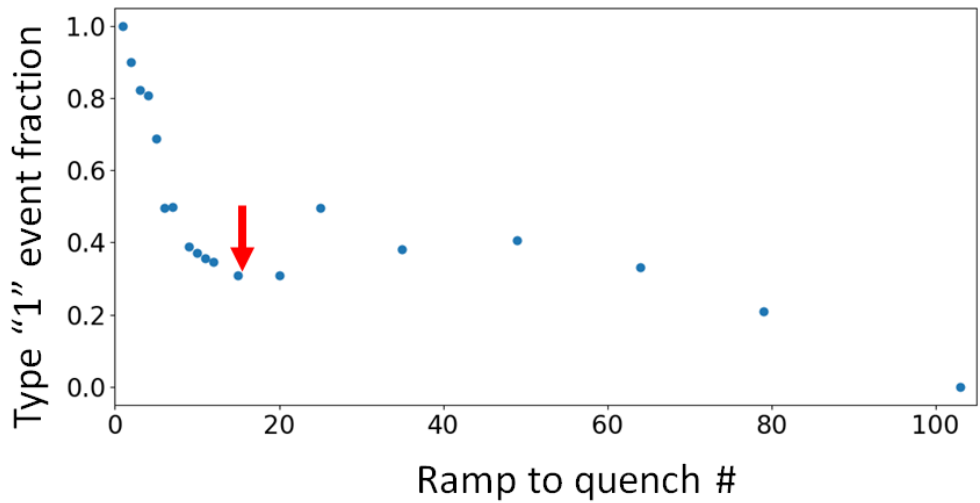

Fig. 5. A plot showing variation of the relative fraction of "type 1" events similar to those in quench #1 as function of the training quench number in CCT4 magnet, as determined using trained random forest classification technique.

**Active acoustic methods**

In addition to passive methods, active techniques have been developed recently that involve sending a pulsed acoustic excitation to the magnet and monitoring its response. Active acoustics allows to:

  iv.   Access interfacial stability during magnet energization and training cycle [23]
  v.    Measure real-time temperature variations in the bulk of the magnet [24]

A technique expanding capabilities of (v) to spatially localize heat sources in the magnet is presently under development at LBNL, and first results have already been demonstrated on conductor samples. Another promising application of active acoustics could be a precise focusing of ultrasonic excitation into a particular target volume within the magnet winding using time-reversal principles [25]. We aim to adapt it for direct measurements of non-elasticity (= dissipation) at a specified location in the bulk of the magnet through second harmonic generation [26]. Hypothetically, same approach can also be used to directly affect magnet behavior by enabling or facilitating local slip stick-motion with ultrasonic vibrations [27] of sufficiently large amplitude. If successful, it would open up a practical way for mitigating premature quenching and accelerating or bypassing the training process altogether without quenching the magnet.

**Conclusion**

We have summarized the current state of the art in AE diagnostics aiming at understanding mechanical transients in superconducting accelerator magnets and their connection to the training process. It is important to stress that diverse AE diagnostics capabilities can be achieved using a unified instrumentation suite that would include an array of cryogenic acoustic sensors mounted along free magnet surfaces, several ultrasonic transducers to provide pulsed and arbitrary waveform ultrasonic excitation, and a specialized software for transducer control and signal analysis. Several key components of such package already exist or under development by the US Magnet Development Program (US MDP). In a short term (2-3 years) we propose to finalize and complete this package, compile a unified diagnostic measurement protocol associated with it, and develop a synergistic analysis approach to correlate acoustic data with the data obtained with other diagnostics such as voltage taps, quench antennas, optical fibers, etc. Then, in a longer time frame (5-10 years) we would systematically deploy our acoustic diagnostic package for accelerator magnet tests conducted by the US MDP, as well as tests conducted for various specialty magnet projects (medical gantries, test facility dipole, fusion magnets) and large international collaboration such as AUP.

This way, a significant amount of unique diagnostic data will be collected, deeper understanding of transient mechanics in magnets under stress will be gained in a systematic way, and practical recommendation will be made for designing training-free accelerator magnets. This effort can be accomplished with a modest investment, and in a way that is minimally intrusive to magnets under test. It will have a significant impact on solving the training problem for the next generation of high-field accelerator magnets in future particle colliders.